\documentclass[fleqn,usenatbib]{mnras}

\usepackage{newtxtext,newtxmath}
\usepackage[T1]{fontenc}

\DeclareRobustCommand{\VAN}[3]{#2}
\let\VANthebibliography\thebibliography
\def\thebibliography{\DeclareRobustCommand{\VAN}[3]{##3}\VANthebibliography}

\usepackage{graphicx}	% Including figure files
\usepackage{amsmath}	% Adhttps://www.overleaf.com/projectvanced maths commands
\usepackage{hyperref}
\usepackage{adjustbox}
\usepackage{stfloats}
\usepackage{graphicx}% Include figure files
\usepackage{dcolumn}% Align table columns on decimal point
\usepackage{bm}% bold math
\usepackage{amsmath}
\usepackage{comment}
\usepackage{booktabs}
\title[Prospects observe SS 433 with LACT]{Prospects for Observing the Microquasar SS 433 with the LACT Array}

\author[Zhen Xie et al.]{
Zhen Xie,$^{1,2,3}$
Zhipeng Zhang,$^{1,2,3}$
and Ruizhi Yang$^{1,2,3}$\thanks{E-mail: yangrz@ustc.edu.cn}
\\
$^{1}$School of Physical Sciences, University of Science and Technology of China, Hefei 230026, China\\
$^{2}$School of Astronomy and Space Science, University of Science and Technology of China, Hefei, Anhui 230026, China\\
$^{3}$Tianfu Cosmic Ray Research Center, Chengdu, China
}

\date{Accepted XXX. Received YYY; in original form ZZZ}

\pubyear{\the\year{}}

\begin{document}
\label{firstpage}
\pagerange{\pageref{firstpage}--\pageref{lastpage}}
\maketitle

\begin{abstract}
We investigate the observational capabilities of the upcoming LACT Cherenkov telescope array for the microquasar SS 433 through detailed simulations. Our results indicate that a detection significance of $5\sigma$ can be achieved with approximately 30 hours of observation; this exposure, coupled with LACT's excellent angular resolution, enables the spatial separation of the eastern and western jets. Furthermore, based on the LHAASO spectral and morphological findings, the array is expected to distinguish the central hadronic component after roughly 100 hours of observation. We also examine its ability to differentiate between the H.E.S.S. and LHAASO spectral models. These findings demonstrate LACT’s strong potential to eventually provide critical insights into particle acceleration in PeVatrons and the radiation mechanisms of microquasars.
\end{abstract}


%%%%%%%%%%%%%%%%%%%%%%%%%%%%%%%%%%%%%%%%%%%%%%%%%%

%%%%%%%%%%%%%%%%% BODY OF PAPER %%%%%%%%%%%%%%%%%%

\section{Introduction}
Microquasars are compact binary systems in our Galaxy, composed of a compact object—either a neutron star or a black hole—accreting matter from a stellar companion(\cite{mirabel2007microquasars}). They are often associated with powerful, mildly relativistic jets that resemble(\cite{bosch2009understanding}), on smaller scales, the relativistic outflows observed in active galactic nuclei (AGN). Owing to their proximity and shorter variability timescales, microquasars serve as unique laboratories for studying jet physics, particle acceleration, and non-thermal emission processes under conditions analogous to those in AGN, but accessible on humanly observable timescales(\cite{belloni2009jet}). The microquasar SS 433 offers a particularly instructive case. Its jets, moving at approximately $0.26c$ (\cite{blundell2004symmetry}), are mildly relativistic and precess with a period of 162 days (e.g. \cite{fabian1979ss,milgrom1979interpretation,margon1979bizarre,hjellming1981analysis}), carrying a kinetic power of about $10^{39}~$erg/s (\cite{margon1982relativistic}). The inner jets, within $\sim$0.1 pc of the central binary (\cite{spencer1979radio,blundell2004symmetry}), have been extensively studied in the radio, optical, and X-ray bands, revealing their precession, collimation, and emission properties. In contrast, the outer jets, extending to $\sim$25 pc, remain less understood. Observationally, they exhibit sudden re-brightening in X-rays and TeV gamma rays, suggesting localized particle acceleration or interactions with the surrounding medium. The contrast between the well-characterized inner jets and the enigmatic outer jets highlights the importance of resolving both spatial and spectral features to understand jet dynamics and radiation mechanisms.

SS 433 also stands out as an important object in the very-high-energy regime. The High-Altitude Water Cherenkov (HAWC) observatory first reported TeV gamma-ray emission from the outer jets(\cite{olivera2023resolving,alfaro2024spectral}), providing compelling evidence for efficient particle acceleration up to multi-TeV energies. More recently, the High Energy Stereoscopic System (H.E.S.S.) confirmed this detection with superior angular and energy resolution, enabling detailed morphological and spectral studies(\cite{hess2024acceleration}). These observations revealed a striking energy-dependent morphology: while TeV photons are observed across extended regions of the jets, photons above $\sim$10 TeV are confined to the vicinity of the jet bases. This result strongly disfavors a purely hadronic origin of the emission and instead points to efficient local particle acceleration, most likely at shock sites, coupled with rapid cooling.

In the ultra-high-energy(E $\geq$ 100 TeV) range,the Large High Altitude Air Shower Observatory (LHAASO)(\cite{zhen2019introduction}) has recently extended the observational window. In this region, LHAASO reported extended ultra high energy (UHE) gamma-ray emission from the central region of the system(\cite{lhaaso2024ultrahigh}), coinciding with a giant gas cloud, suggesting a possible hadronic origin for at least part of the emission. Beyond SS 433, LHAASO has also detected UHE emission from several other black hole X-ray binaries, with spectra in some cases extending up to hundreds of TeV. These findings demonstrate that accreting black holes and their environments can act as extremely efficient cosmic accelerators, potentially contributing to Galactic cosmic rays in the PeV regime around the so-called “knee” of the spectrum. These observations establish SS 433 as a prototypical system for investigating jet-powered acceleration processes in compact binaries.

Yet, many key questions remain unresolved, including the physical origin of the jet re-brightening at large distances, the dominant radiative processes at work, and the dynamical properties of the outer jets. Addressing these open issues requires next-generation instruments that combine high sensitivity at multi-TeV energies with wide-field coverage and excellent angular resolution.

The Large Array of imaging atmospheric Cherenkov Telescopes (LACT; \cite{zhang2024layout,zhang2025performance}), currently under construction at the same site as the LHAASO experiment, is designed to meet these requirements. LACT will naturally complement the wide energy coverage of the LHAASO air-shower array with its superior angular and energy resolutions at multi-TeV energies.

In particular, recent LHAASO observations have been interpreted as evidence for a hadronic component in the gamma-ray emission (\cite{lhaaso2024ultrahigh}), possibly arising from relativistic protons escaping from the jets and interacting with surrounding molecular clouds. However, the limited angular resolution of LHAASO hinders the detailed investigation of complex regions like the SS 433 jets and their surroundings. Furthermore, the potentially comparable fluxes of the leptonic and hadronic components make it even more challenging to disentangle their respective contributions without superior sensitivity and resolution. While H.E.S.S. provides sufficient angular resolution for morphological studies, its limited sensitivity at energies of several tens of TeV restricts a systematic exploration of this critical energy range.

The tens-of-TeV band is critical for identifying potential spectral or spatial transitions between leptonic and hadronic emission components. While recent observations by HAWC, H.E.S.S., and LHAASO have made significant progress, a systematic exploration of this regime remains limited. Specifically, LHAASO’s angular resolution is insufficient to resolve the fine morphology of extended structures like the SS 433 jets, while H.E.S.S. lacks the sensitivity required at several tens of TeV. Furthermore, since the fluxes of the two emission components may be comparable, disentangling them requires a precision that current instruments cannot fully provide.

From a broader perspective, future LACT observations of SS 433 have the potential to address several fundamental questions in high-energy astrophysics. Owing to its excellent angular resolution and sensitivity at multi-TeV energies, LACT can measure the energy-dependent morphology and spectra of the jets, providing direct tests of particle acceleration and transport scenarios. In particular, resolving the spatial distribution of gamma-ray emission in the tens-of-TeV range—where current observations remain limited—will place critical constraints on proton escape, diffusion, and interactions with the surrounding environment, and help distinguish between leptonic and hadronic emission mechanisms. More generally, characterizing particle acceleration in SS 433 will offer valuable insights into the physics of mildly relativistic jets and clarify their potential role as contributors to the Galactic high-energy particle population.

The paper is structured as follows. This Section reviews the current understanding of SS 433, with particular emphasis on its high-energy emission mechanisms and the open problems highlighted by recent observations. Section 2 introduces the LACT array, outlining its design features and performance advantages. Section 3 presents preliminary simulations and analysis strategies to evaluate the capability of LACT to detect and resolve the emission from SS 433. Finally, the last Section summarizes the key findings and discusses their implications for future observations of microquasars and Galactic particle acceleration.

\section{Overview of the LACT array}
LACT is designed as a next-generation facility capable of addressing the limitations of existing ground-based gamma-ray observatories. The array consists of 32 telescopes, each with a 6-meter diameter, and each telescope is equipped with a camera based on silicon photomultiplier (SiPM) technology, a well-validated solution previously demonstrated on LHAASO-WFCTA(\cite{aharonian2021construction}). The use of SiPM cameras enables operation during moonlit nights, significantly increasing the effective observation time. LACT is optimized to combine high angular resolution with broad energy coverage, providing sub-degree imaging of extended sources such as SS 433’s jets, while maintaining sensitivity down to energies below 1 TeV to allow observations of extragalactic sources and gamma-ray transients detected by LHAASO-WCDA(\cite{hu2023first}).

Unlike extensive air-shower arrays such as LHAASO, imaging atmospheric Cherenkov telescopes (IACTs) operate only during moonless nights, resulting in a much lower duty cycle. To compensate for the reduced observation time and detect a comparable number of ultra-high-energy gamma-ray events, LACT is designed with an effective area significantly larger than that of current-generation IACTs, exceeding 1 $\rm km^2$. In its standard observation mode, the full array operates as a unified system, providing excellent performance across a wide energy range, from a few hundred GeV up to hundreds of TeV. This mode is well suited not only for detecting ultra-high-energy gamma rays, but also for observing variable and transient sources such as active galactic nuclei and gamma-ray bursts.

To further enhance sensitivity to the highest-energy events, LACT supports a large-zenith-angle (LZA) observation mode, a technique widely applied in existing IACTs (for example, see \cite{konopelko1999effectiveness,acciari2020magic,adams2021veritas}). Observing at large zenith angles increases the effective area and sky coverage, and at the LHAASO site ($29^\circ21'27.6''$N, $100^\circ08'19.6''$E), it allows access to regions such as the Galactic Center, which can only be observed above zenith angles of 50°. In this mode, the 32 telescopes are divided into eight cells, each containing four closely spaced telescopes, and telescopes from different cells can be combined to form four groups. At the high altitude of the LHAASO site (4400 m), LZA observations are particularly beneficial: the increased atmospheric depth between the shower maximum and the telescopes flattens the lateral distribution of Cherenkov photons, producing smaller and higher-quality images and mitigating potential image leakage \cite{zhang2024layout}. And by organizing the array into four sub-arrays, this approach can potentially extend the effective observation time by up to a factor of four. This configuration is therefore optimized for detecting ultra-high-energy gamma-ray events and complements the standard observation mode by extending the array’s sensitivity to the highest energies.

\begin{figure}
    \centering
    \includegraphics[width=0.85\linewidth]{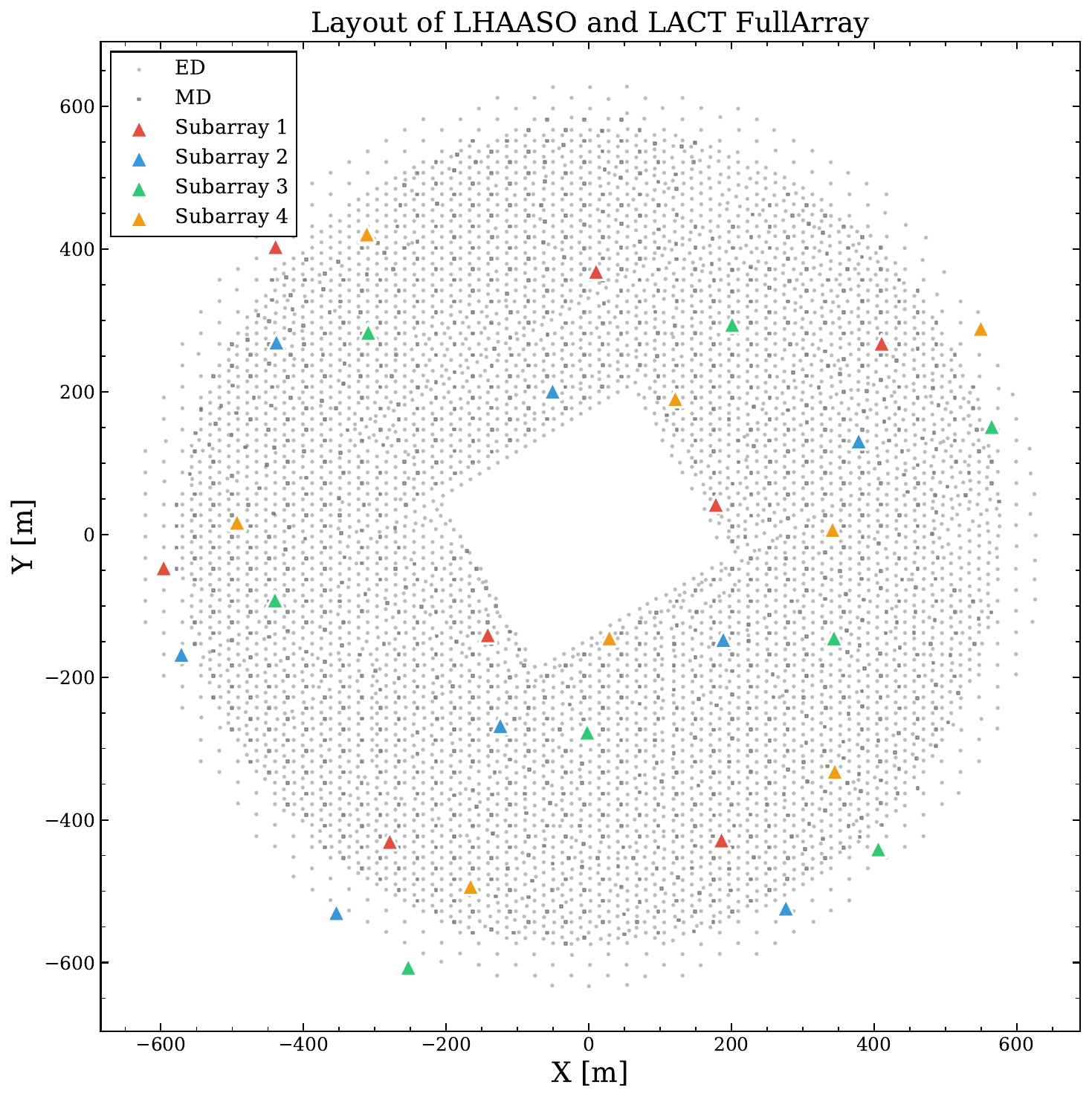}
    \includegraphics[width=1\linewidth]{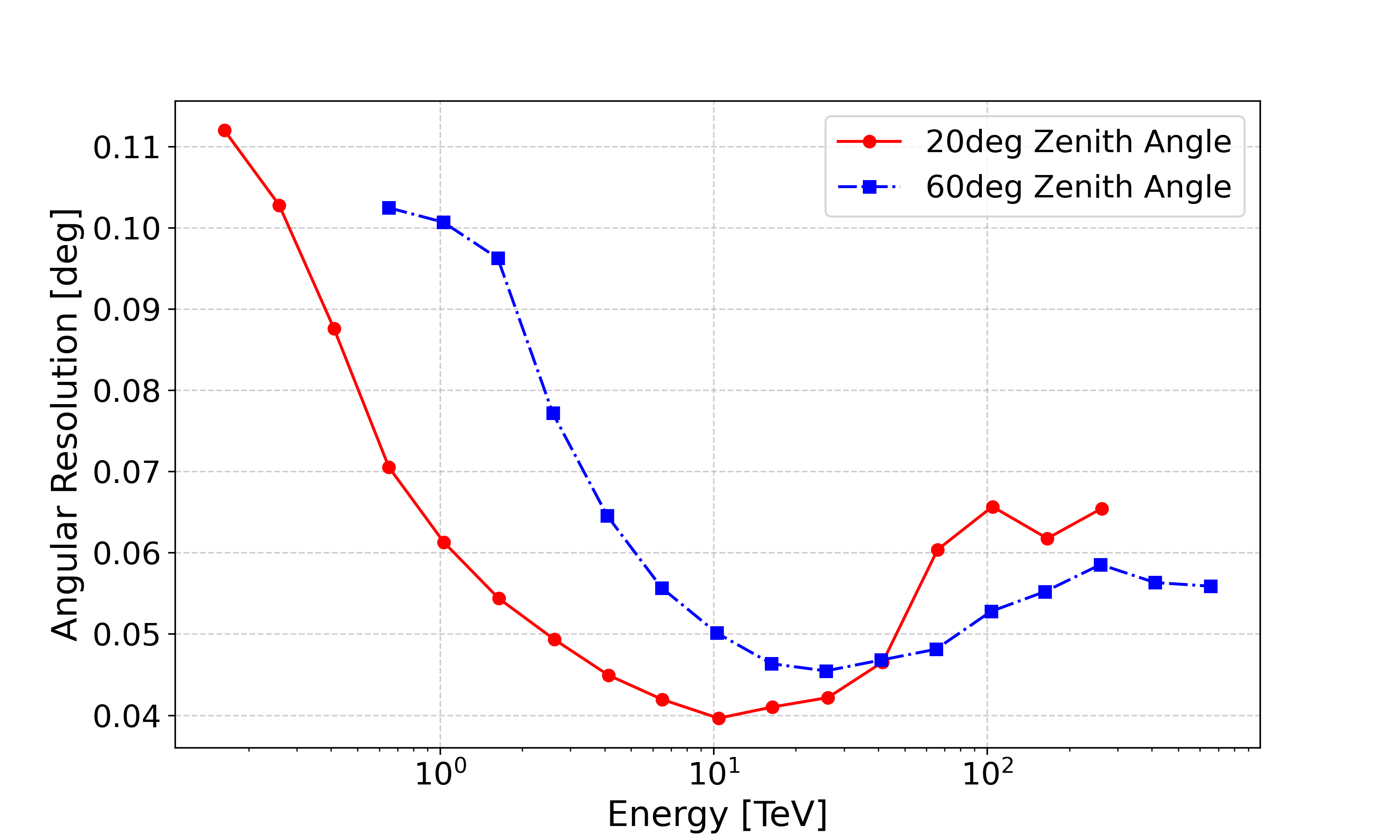}
    \includegraphics[width=0.85\linewidth]{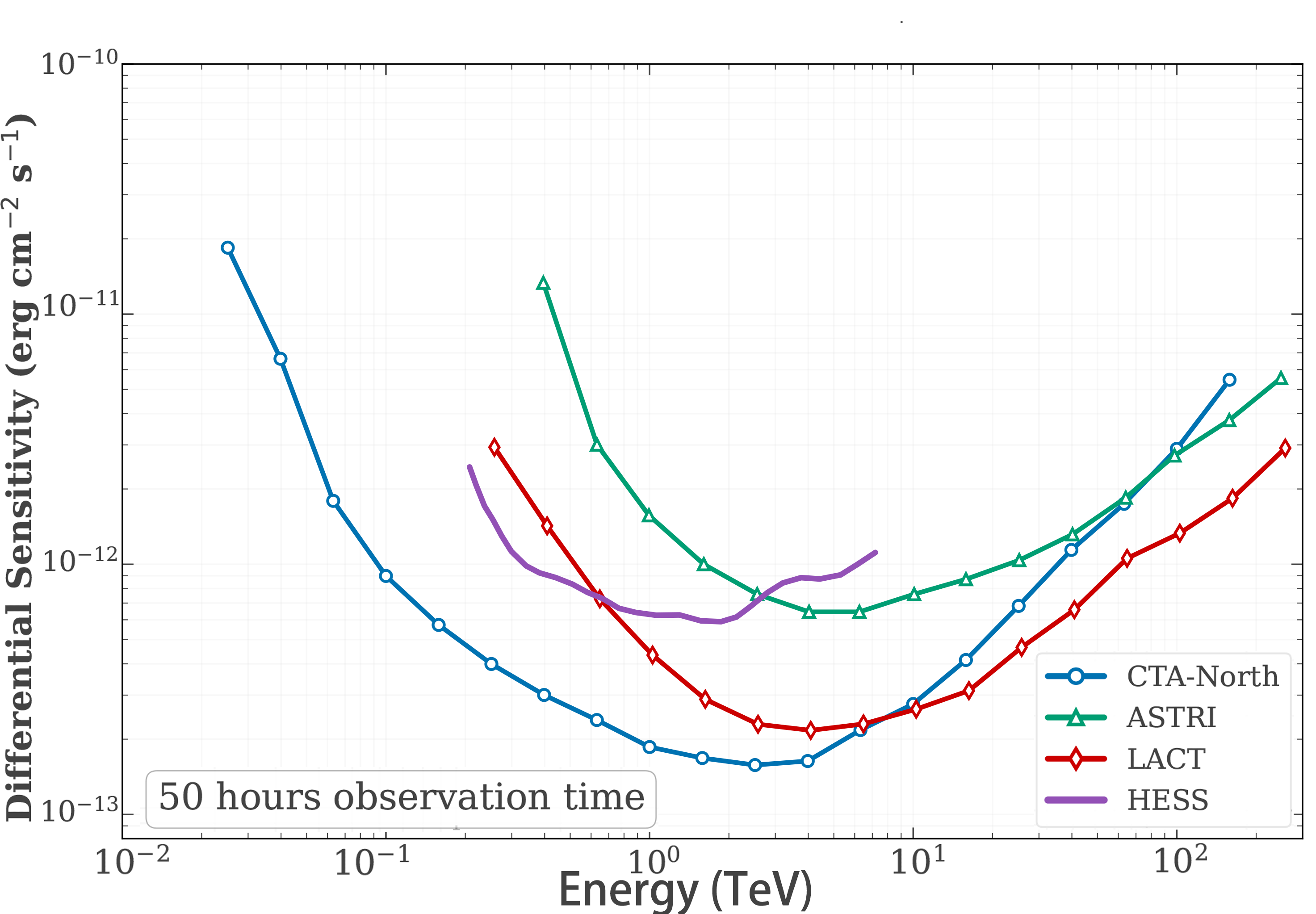}

    \caption[Description of the optimized array layout.]
    {Top: The placement of the telescopes within the LACT array.
    Middle: The angular resolution as a function of energy for two zenith angle configurations(\cite{zhang2025performance}).  
    Bottom: A comparison of the LACT sensitivity (at $20^\circ$ zenith angle mode) against current generation IACTs—H.E.S.S. (\cite{aharonian2006observations}), ASTRI (\cite{pareschi2016astri})—as well as the future CTAO North(\cite{acharya2017science}). 
    }
    \label{fig:effa}
\end{figure}

In previous work by \cite{zhang2025performance}, detailed Monte Carlo simulations were carried out to determine the optimized LACT layout and evaluate the instrument performance under two zenith angle configurations. For the present study, we adopt the instrument response functions corresponding to these two modes, with zenith angles of 20° and 60°, and the resulting angular resolution and effective area estimates are shown in Fig.~\ref{fig:effa}.

\section{Evaluation of Observational Capabilities}
To assess LACT’s observational capabilities for the microquasar SS~433, we performed detailed simulations based on the instrument response functions (IRFs) derived from the layout in\cite{zhang2024layout}. These IRFs incorporate key performance characteristics, including effective area, angular resolution, energy dispersion, and background rates (see Fig.~\ref{fig:effa}). Simulations were carried out for the two zenith angle configurations adopted in this study, 20° and 60°, representing near-zenith and large-zenith-angle observations, respectively.

\begin{figure}
    \centering
    \includegraphics[width=1.0\linewidth]{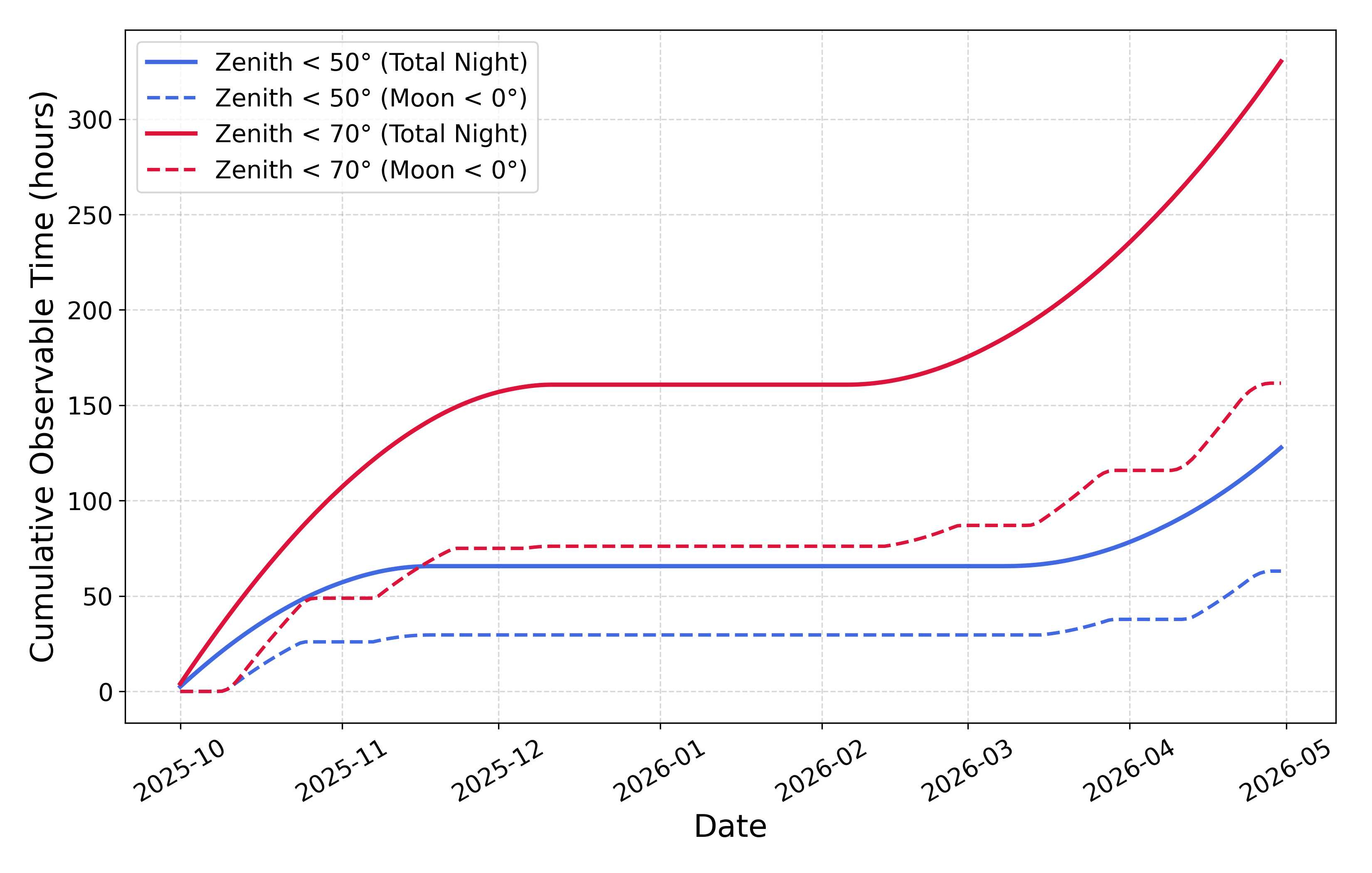}
    \caption{Cumulative observable time of SS~433 from the LHAASO site between October and April. The curves show the total visibility for zenith angles below 50° and below 70°, with dashed lines indicating the corresponding observation time restricted to moonless nights. Nighttime is defined as periods when the Sun’s altitude is below –18° (Astronomical Twilight).}
    \label{fig:ot}
\end{figure}

Based on the source visibility from the LHAASO site, we estimated the effective observation time for SS~433 over the typical observing season, from October to April, under favorable weather conditions. To represent the two observation modes, we considered two zenith angle ranges: 0°–50° for the near-zenith configuration and 0°–70° for the large-zenith-angle configuration. Nighttime was defined as the period when the Sun’s altitude is below –18° (Astronomical Twilight), and the target’s altitude was computed using the \texttt{Astropy} framework (\cite{The_Astropy_Collaboration_2022}). The cumulative observable hours for the two zenith angle ranges are shown in Fig.~\ref{fig:ot}. Over the six-month season, the effective observation time for zenith angles below 50° reaches a total of 127.8 hours, and 63.1 hours when limited to moonless nights. For zenith angles below 70°, the total observation time increases to 330.3 hours, with 161.6 hours available on moonless nights. These results demonstrate that SS~433 can be effectively observed from the LHAASO site.

A central aspect of the analysis is the study of the detection significance as a function of observation time. For each simulated dataset, significance maps were generated to evaluate the confidence level of source detection in both the eastern and western jets. To provide a realistic template for the simulations, the spatial and spectral properties of the jets were modeled based on H.E.S.S. observations, ensuring that the synthetic datasets reflect the best current knowledge of SS 433, which shown in Table.\ref{tab:combined_info_deg}. By tracking the accumulation of significance with increasing exposure, we can estimate the minimum observation time required for a robust detection of extended structures and localized high-energy features. This approach also allows LACT to leverage its excellent spatial resolution to assess the gamma-ray source's structure, providing a means to distinguish between emission from compact acceleration regions near the jet bases and more extended radiation along the jet flow. Additionally, the nearby source HESS J1908+063 and the background model were included in the simulations to ensure a more realistic representation of the observational environment.

\begin{table}
\centering
\caption{Basic Information for jets of SS 433 from H.E.S.S observation spatial model above 10 TeV, modeled with a symmetric Gaussian. $E_0$ is the reference energy, $\phi_0$ is the amplitude at the reference energy, and $\Gamma$ is the photon spectral index. $l$ and $b$ are the Galactic longitude and latitude of the central position. Physical sizes for $r$ are calculated assuming a distance of 5.5 kpc. The unit of $\phi_0$ is $10^{-13}$ TeV$^{-1}$ cm$^{-2}$ s$^{-1}$.}
\label{tab:combined_info_deg}
\begin{tabular}{l c c}
\toprule
Parameter & east & west \\
\midrule
$l$ (deg) & $39.840 ^{\pm 0.031}$ & $39.560 ^{\pm 0.010}$ \\
$b$ (deg) & $-2.643 ^{\pm 0.038}$ & $-1.951^{\pm 0.011}$ \\
$r$ (deg) & $0.013 ^{\pm 0.029}$ & - \\
$\phi_0$ & $2.30 ^{\pm 0.58_{\text{stat.}}\pm 0.32_{\text{syst.}}}$ & $2.83 ^{\pm 0.70_{\text{stat.}}\pm 0.39{\text{syst.}}}$ \\
$\Gamma$ & $2.19 ^{\pm 0.12_{\text{stat.}} \pm 0.12_{\text{syst.}}}$ & $2.40 ^{\pm 0.15_{\text{stat.}} \pm 0.13_{\text{syst.}}}$ \\
\bottomrule
\end{tabular}
\end{table}
\begin{table}
\centering
\caption{Basic properties of the source LHAASO J1911+0513, which observed by LHAASO in SS 433 central region at energies above 100 TeV. The table lists the original equatorial coordinates (RA, Dec) and the 39\% containment radius $r_{39}$, as well as the corresponding galactic coordinates (l, b) of the source center. At 100 TeV, 1 CU = $6.1 \times 10^{-17}~\rm TeV^{-1} cm^{-2} s^{-1}$.}
\label{tab:lhaaso}
\begin{tabular}{l c c c}
\toprule
 coordinate&position(deg)  &$r_{39}$(deg)& flux \\
\midrule
 (RA,Dec)&($287.89 ^{\pm 0.11}$, $5.22^{ \pm 0.12}$) &$0.32^{\pm 0.09}$& 0.32 CU\\
 (l,b)&(39.87, -2.08)& &(in 100 TeV) \\
\bottomrule
\end{tabular}
\end{table}

The simulations were carried out using the \texttt{Gammapy-1.3}(\cite{donath2023gammapy}) analysis framework, which was extended to handle the IRFs specific to LACT. By combining the source model with the instrument response, we generated mock event lists and reconstructed both the spatial and spectral characteristics of the jets, which energy band is 10 - 300 TeV. Standard analysis procedures, including background estimation, and likelihood fitting, were applied to evaluate the detectability of the outer jets and the accuracy of the spectral reconstruction. The use of \texttt{Gammapy} enables a flexible and well-tested platform for this study, providing robust estimates of significance, flux, and morphology, while also allowing for direct comparison with existing H.E.S.S. results.
To ensure that the results are statistically robust and not affected by random fluctuations, we performed 100 independent realizations for each observation time. For each simulation, the significance of the detection was calculated separately for the eastern and western jets of SS 433, allowing us to evaluate the expected detection confidence as a function of exposure. By aggregating the results from these repeated simulations, we were able to determine the mean significance and the corresponding $1\sigma$ uncertainty (68\% confidence interval) for each jet at different observation durations.

Preliminary simulations indicate that LACT can achieve high-confidence detection of the outer jets of SS 433 with relatively modest observation times. Specifically, a detection significance exceeding $3\sigma$ is expected after approximately 20 hours of exposure. In comparison, our current observations (which utilized 200 hours of exposure) have yielded significances of roughly $13.79\sigma$ and $11.25\sigma$ for the east and west jets, respectively, exceeding the roughly $7.8\sigma$ and $6.8\sigma$ reported by prior H.E.S.S. observations of similar duration(\cite{hess2024acceleration}). Representative results of these simulations, including the evolution of detection significance for the east and west jets, are presented in Figure.\ref{fig:svt}. These results demonstrate that LACT can achieve high-confidence detection for both the eastern and western jets within reasonable observation times, while also providing good spatial resolution for a clear observation of both jets.
\begin{figure}
    \centering
    \includegraphics[width=0.95\linewidth]{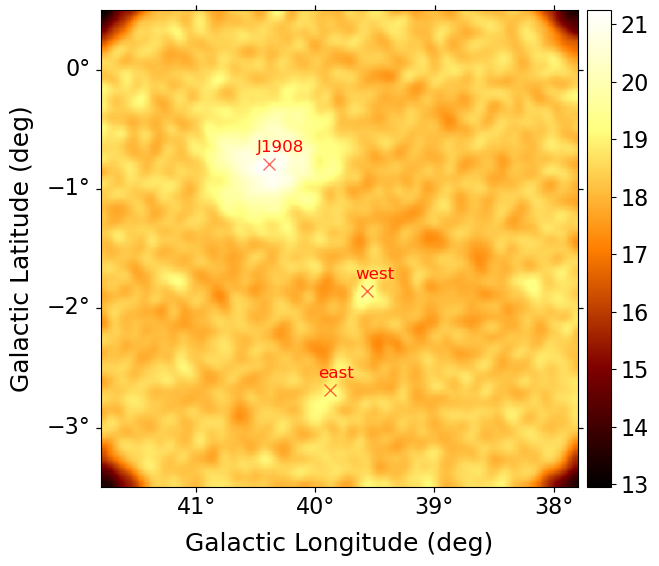}
    \includegraphics[width=0.95\linewidth]{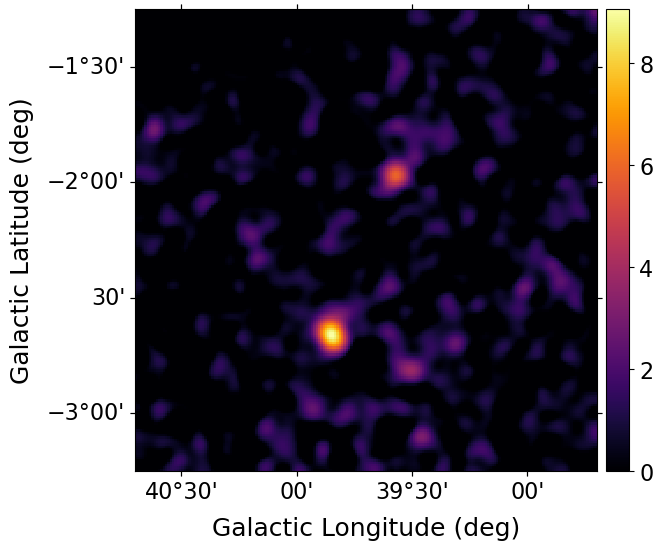}
    \caption{Top: A counts map of the SS 433 eastern and western jets from an 100-hour simulated observation. Bottom: The corresponding significance map in SS 433 region, where the significance is represented as the square root of the TS value.}
    \label{fig:sigmap}
\end{figure}
\begin{figure}
    \centering
    \includegraphics[width=1\linewidth]{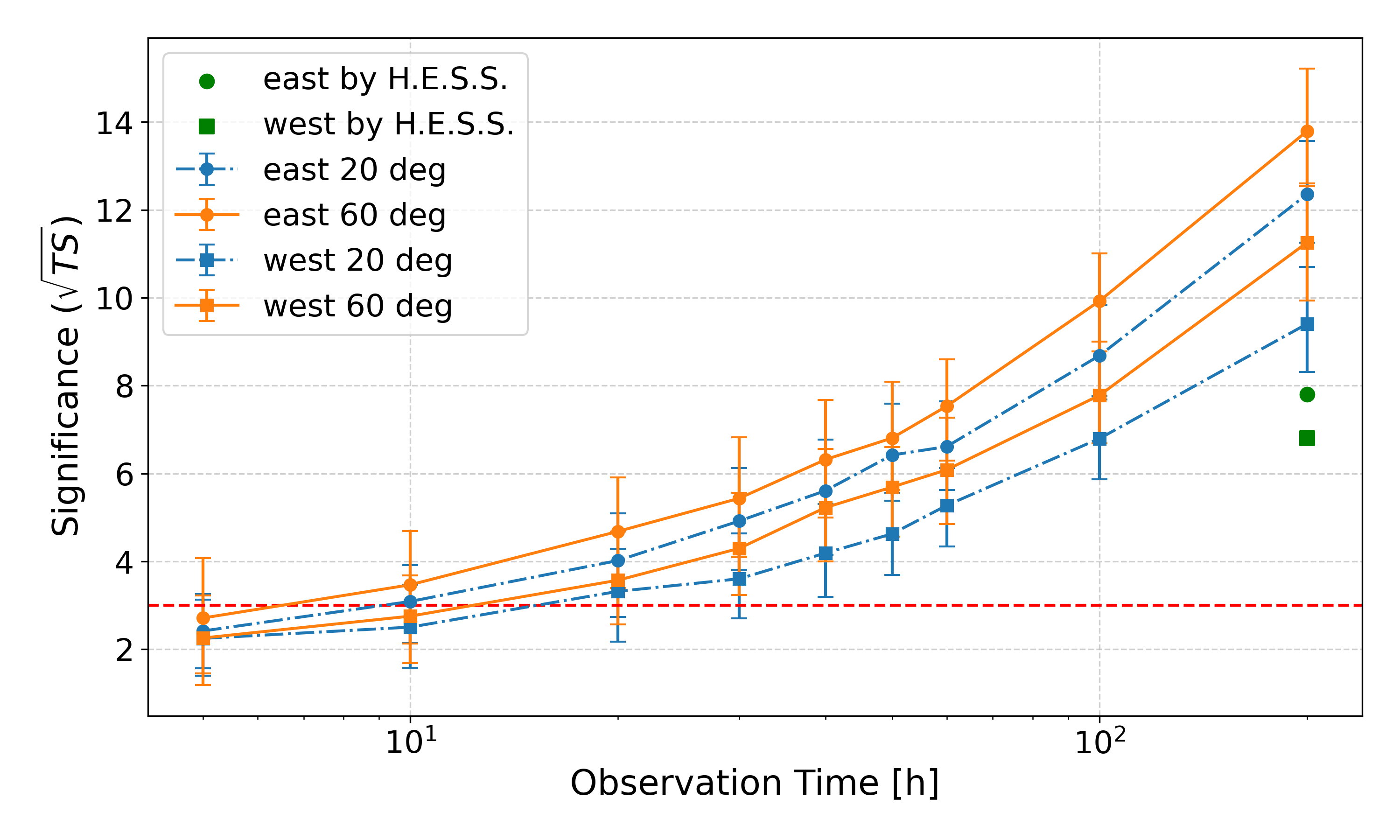}
    \caption{The detection significance of the SS~433 eastern and western jets as a function of observation time, integrated over all energies from 10 to 300 TeV}. The red dashed line indicates the $3\sigma$ threshold. Each data point represents the mean of total simulations, with error bars corresponding to the 68\% confidence interval. The green dots represents the result of H.E.S.S..
    \label{fig:svt}
\end{figure}
\begin{figure}
    \centering
    \includegraphics[width=1\linewidth]{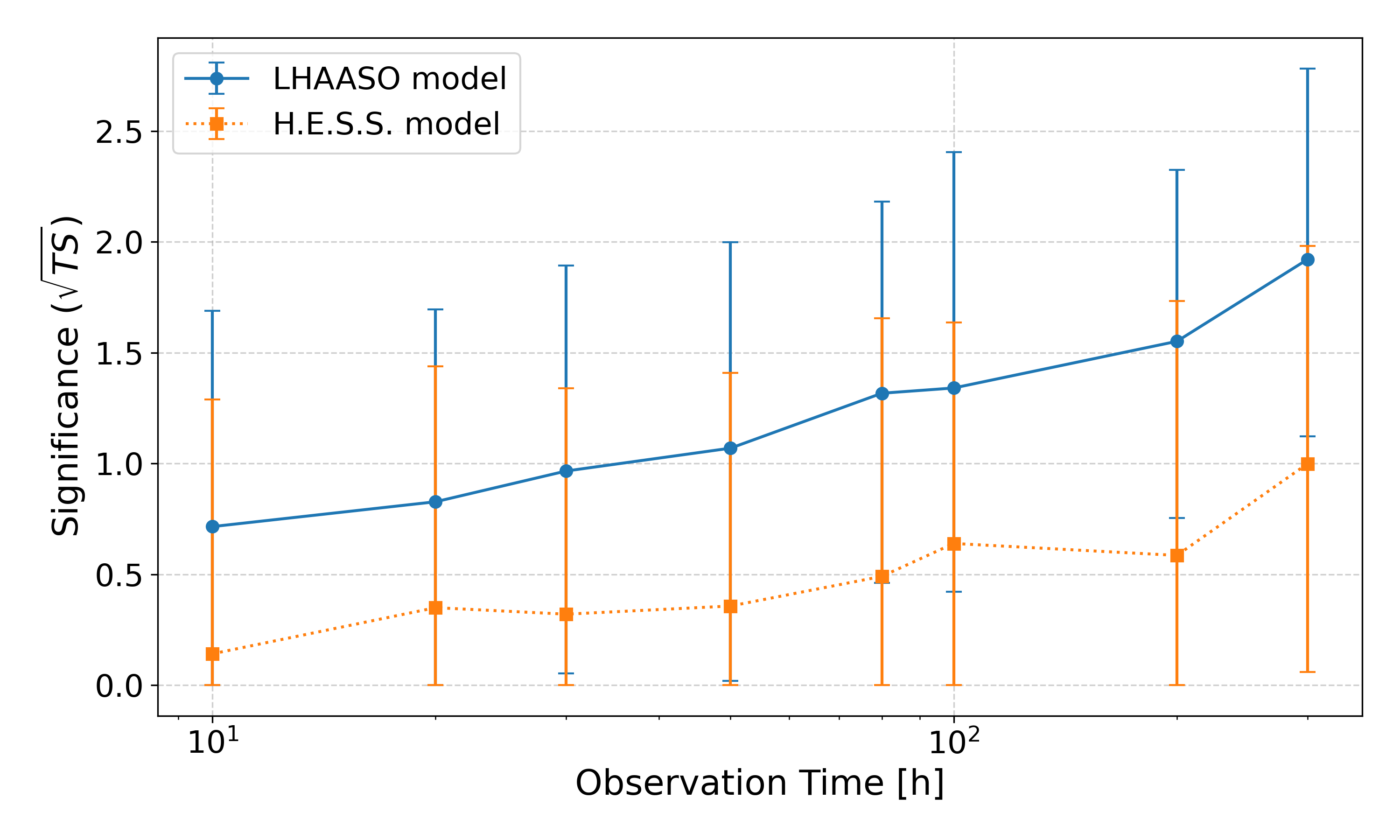}
    \caption{Difference in significance between fits with H.E.S.S. and LHAASO templates for events above 100 TeV, with the mean significance as a function of observation time. Each data point represents the mean of total simulations, performed in the $60^{\circ}$ zenith-angle mode. }
    \label{fig:delta_ts}
\end{figure}

Another important observations of SS 433 is that LHAASO's results above 100 TeV point towards an extended source LHAASO J1911+0513(\cite{lhaaso2024ultrahigh}), the basic information is listed in Table.\ref{tab:lhaaso}), and the spectrum of the extended source above $100~\mathrm{TeV}$ is
\begin{align*}
\frac{dN}{dE} = (3.45 \pm 1.61) \times 10^{-16}
\left( \frac{E}{50~\mathrm{TeV}} \right)&^{-3.96 \pm 0.25}
&~\mathrm{TeV^{-1}\,cm^{-2}\,s^{-1}}
\label{eq:spectrum_extended}
\end{align*}

To investigate the capability of LACT to distinguish different emission scenarios of SS~433, we adopted a consistent analysis framework based on the three-source model inferred from the LHAASO observation (\cite{lhaaso2024ultrahigh}). Using this model as the input source template, we simulated LACT observations at a zenith angle of $60^{\circ}$ with the corresponding IRFs, and fitted the simulated data with both the LHAASO and H.E.S.S. spectral models to assess their distinguishability.

We first focused on photons with energies above 100~TeV to test LACT’s performance in the ultra-high-energy regime. The difference in detection significance between the two models is shown in Fig.~\ref{fig:delta_ts}. As expected, the detection significance gradually increases with exposure time. However, it should be noted that while H.E.S.S. observations have not yet clearly resolved a spectral cutoff, a suppression of the flux at the highest energies is physically expected due to the Klein-Nishina effect on the electron distribution. Consequently, the significance presented here might be somewhat overestimated, as the flux above 100~TeV could be even lower than the pure power-law extrapolation, making it difficult to reach statistical significance within a reasonable observation time.

Considering the LHAASO results (Fig.~1e in \cite{lhaaso2024ultrahigh}), which suggest that the hadronic-origin emission still makes a substantial contribution below 100~TeV, we extended the spectral range of the LHAASO-like extended source by including this additional hadronic component. In the simulation, we constructed a three-component model consisting of the central hadronic component and the two leptonic jet components, following the spatial and spectral characteristics inferred from LHAASO observations. The same three-component configuration was adopted in the fitting procedure to evaluate how well LACT can recover the central hadronic emission under different energy thresholds and observation times.

\begin{figure}
    \centering
    \includegraphics[width=1\linewidth]{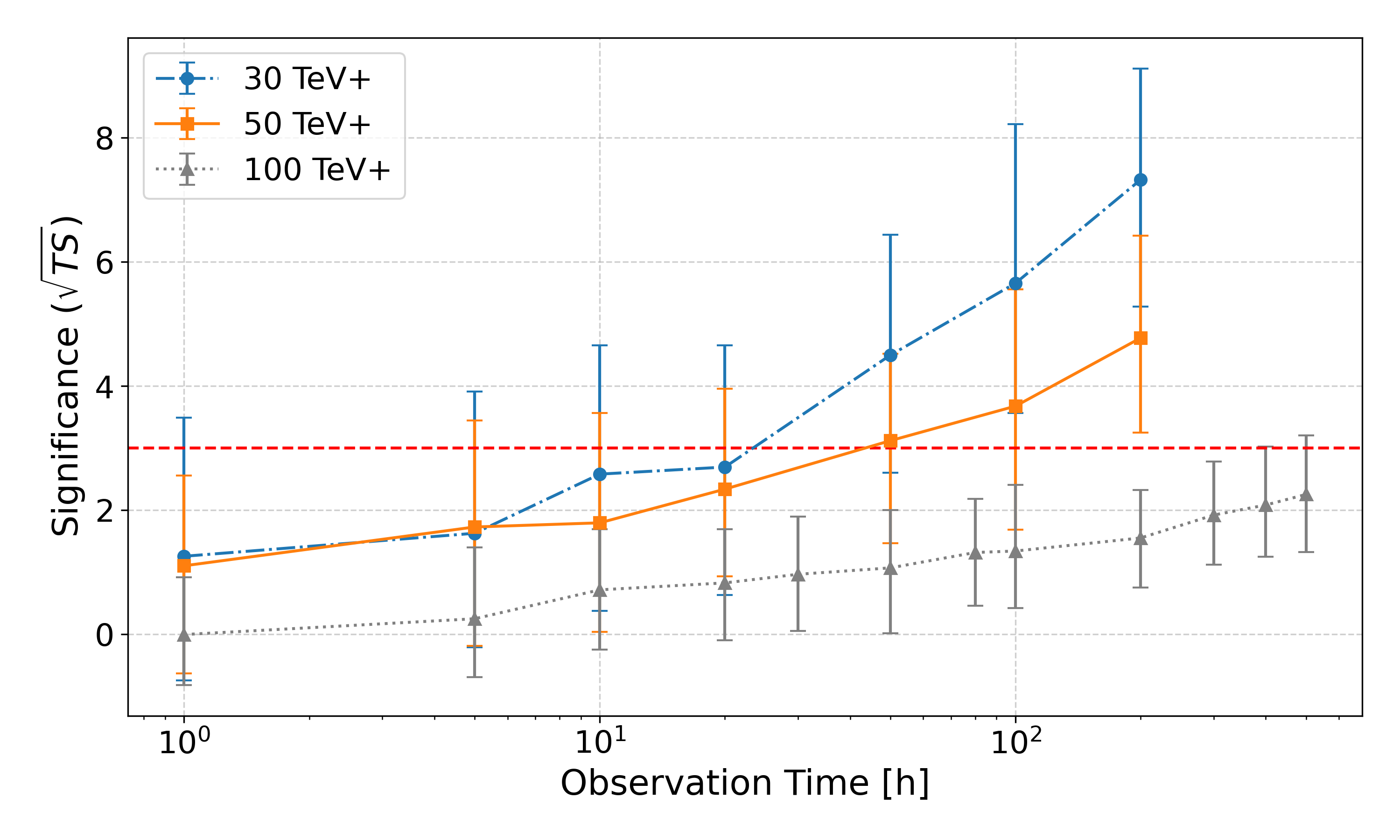}
    \caption{Detection significance of the central hadronic component as a function of observation time for different energy thresholds. The curves correspond to simulated LACT observations above 30~TeV, 50~TeV, and 100~TeV, respectively.}
    \label{fig:hard}
\end{figure}

\begin{figure}
    \centering
    \includegraphics[width=0.95\linewidth]{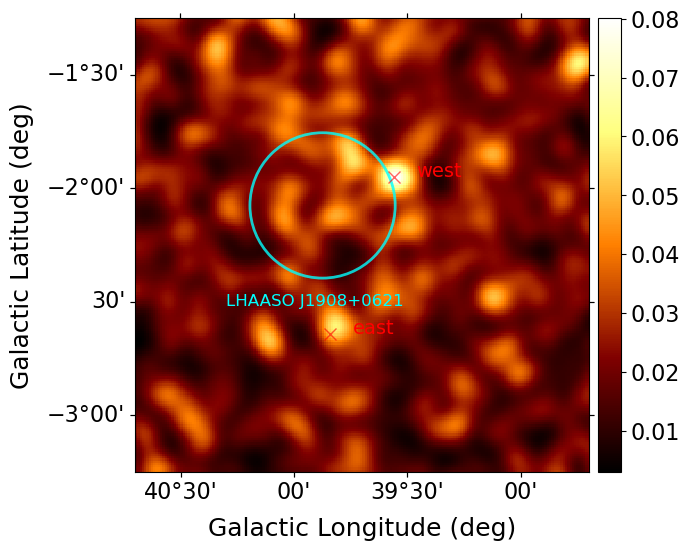}
    \includegraphics[width=0.95\linewidth]{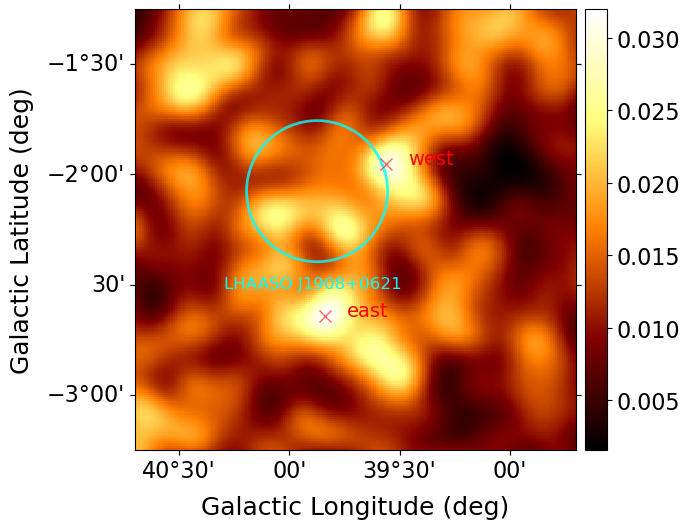}
    \includegraphics[width=0.95\linewidth]{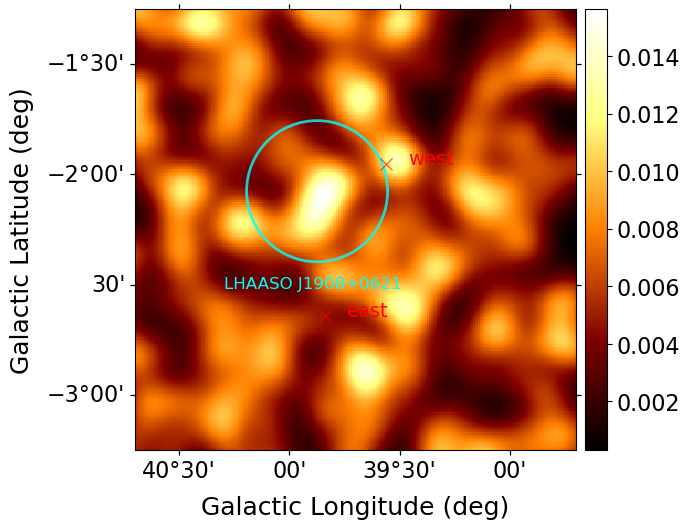}
    \caption{Simulated LACT counts maps for SS~433 based on the three-source model inferred from LHAASO observations. From top to bottom: photon energies above 30~TeV, 50~TeV, and 100~TeV. The red crosses mark the positions of the two jet regions, while the cyan circle indicates the location and approximate extent of the extended source reported by LHAASO. All maps correspond to one 100-hour observation in the 60°-zenith-angle mode using the same IRFs. At energies above 100~TeV, the very low photon counts lead to noticeable irregular fluctuations dominated by the background.}
    \label{fig:cm-3s}
\end{figure}

As shown in Fig.~\ref{fig:hard}, for energies above 30~TeV, LACT would be able to marginally detect (3$\sigma$ significance) the central hadronic component after about 50 h of observation. When the threshold is increased to 50~TeV, achieving a similar significance would require approximately 100 hours of exposure. At higher energies (e.g., above 100~TeV), although the hadronic component becomes increasingly dominant relative to the leptonic emission, as shown in Fig.~\ref{fig:cm-3s}, the photon statistics are too limited for a significant detection. These results indicate that LACT’s excellent angular resolution enables it to start resolving the hadronic-origin emission in the tens-of-TeV range within a feasible observation time.

At energies above $100$ TeV, where photon statistics become increasingly limited for LACT alone, a promising way forward is the joint use of LACT and the LHAASO-KM2A array. By performing a joint reconstruction that leverages the Cherenkov-light information from LACT alongside the particle measurements from KM2A—specifically utilizing the Muon Detectors to achieve near background-free event selection—it becomes possible to significantly increase the available statistics at small zenith angles. This synergy enhances the localization and characterization of ultra-high-energy gamma-ray events, providing more reliable spectral and morphological studies above $100$ TeV. Consequently, such a combined analysis offers much stronger constraints on particle acceleration and transport processes within the SS 433 region.
\section{conclusion}
In this work, we have explored the scientific potential of the LACT for investigating the microquasar SS 433, with particular emphasis on its extended outer jets and their high-energy non-thermal emission. SS 433 represents one of the most remarkable laboratories for studying relativistic jet physics within our Galaxy, providing a unique opportunity to probe particle acceleration and radiation processes under extreme conditions. Although instruments such as H.E.S.S., HAWC, and LHAASO have provided crucial evidence for very-high- and ultra-high-energy $\gamma$-ray emission from this system, their limitations in angular resolution, sensitivity, or energy coverage still prevent a detailed morphological and spectral dissection of the jet regions.

Based on the source visibility calculations, we estimate that at the LHAASO site, during the favorable observing season from October to April, the available observation time amounts to about 120 hours for small zenith angles (0°–50°) and more than 200 hours for large zenith angles (50°–70°), as shown in Fig.\ref{fig:ot}. Our simulation results show that LACT can effectively resolve the eastern and western jet structures of SS~433, achieving a detection significance of approximately 5$\sigma$ within 30 hours in the 10–300~TeV range, with significantly higher sensitivity than current H.E.S.S. observations, which shown in Fig.\ref{fig:svt}. At energies above 100~TeV, LACT is capable of detecting the extended emission reported by LHAASO, but due to the steep spectral decline, substantially longer exposures are required to reach comparable significance(see Fig.\ref{fig:delta_ts}). If SS~433 indeed contains a hadronic-origin component, as suggested by the extended emission observed by LHAASO, LACT would be capable of detecting the corresponding gamma-ray emission from the central region above $30~\rm TeV$, reaching a marginal significance of about $3\sigma$ after $50$~h of observation and about $100$~h above $50~\rm TeV$ as shown in Fig.\ref{fig:hard}. This demonstrates that LACT can start to resolve the hadronic-origin emission within a realistic observation time, even though higher-energy detections are limited by photon statistics.

Overall, our results demonstrate that LACT will be a powerful instrument for high-resolution studies of microquasar jets and other extended $\gamma$-ray sources. The simulations presented in this work establish a quantitative basis for future observational campaigns, providing guidance for the optimization of exposure strategies and observation modes in different zenith-angle configurations. By resolving energy-dependent morphological features and distinguishing localized acceleration zones, LACT can offer new insights into particle transport, cooling, and emission processes that are currently beyond the reach of existing instruments. Beyond SS~433, the simulation framework and analysis approach developed here can be readily applied to a wide range of Galactic and extragalactic $\gamma$-ray sources, opening new opportunities for investigating particle acceleration and non-thermal phenomena in extreme astrophysical environments.
\section*{Acknowledgment}
Rui-zhi Yang is supported by the NSFC under grant 12393854,12588101, and by the natural science funding of Sichuan Province under grant 2025ZNSFSC0065. Rui-zhi Yang gratefully acknowledge the support of Cyrus Chun Ying Tang Foundations and of the studio of Academician Zhao Zhengguo, Deep Space Exploration Laboratory.

We would also like to thank Yanhong Yu and Wenyu Cao for their valuable assistance and insightful discussions regarding the LHAASO data analysis and science related to SS 433.
We also thank the referees for their valuable comments and suggestions, which have greatly helped improve the clarity and presentation of this work.
\section*{Data Availability}
The data presented in this article are simulated event datasets generated within the \texttt{Gammapy} framework, utilizing the Instrument Response Functions (IRFs) derived from LACT. These IRFs were generated based on previous research(\cite{zhang2024layout,zhang2025performance}, and are currently not publicly available.



\bibliographystyle{mnras}
\bibliography{main} % if your bibtex file is called example.bib

%\begin{thebibliography}{99}
%\end{thebibliography}

%%%%%%%%%%%%%%%%%%%%%%%%%%%%%%%%%%%%%%%%%%%%%%%%%%

%%%%%%%%%%%%%%%%% APPENDICES %%%%%%%%%%%%%%%%%%%%%

%%%%%%%%%%%%%%%%%%%%%%%%%%%%%%%%%%%%%%%%%%%%%%%%%%

% Don't change these lines
\bsp	% typesetting comment
\label{lastpage}
\end{document}